\documentclass[11pt]{article}

\usepackage{amsmath}
\usepackage{amssymb}
\usepackage{amsfonts}
\usepackage{latexsym}

\catcode `\@=11
\@addtoreset{equation}{section}

\catcode `\@=12

\newcommand{\be}{\begin{equation}}
\newcommand{\en}{\end{equation}}
\newcommand{\bea}{\begin{eqnarray}}
\newcommand{\ena}{\end{eqnarray}}
\newcommand{\beano}{\begin{eqnarray*}}
\newcommand{\enano}{\end{eqnarray*}}
\newcommand{\bee}{\begin{enumerate}}
\newcommand{\ene}{\end{enumerate}}
\newcommand{\D}{{\cal D}}

\begin{document}

\thispagestyle{empty}

\vspace*{1cm}

\begin{center}
{\Large \bf Multiplication of distributions in any dimension:
applications to $\delta$-function and its derivatives}
\vspace{2cm}\\

{\large F. Bagarello}
\vspace{3mm}\\
  Dipartimento di Metodi e Modelli Matematici,
Facolt\`a di Ingegneria, Universit\`a di Palermo, I - 90128  Palermo, Italy\\
E-mail: bagarell@unipa.it\\web page: www.unipa.it/$^\sim$bagarell
\vspace{2mm}\\
\end{center}

\vspace*{2cm}

\begin{abstract}
\noindent
   In two previous papers the author introduced a multiplication of distributions in one dimension and
   he  proved that two one-dimensional Dirac delta functions and their derivatives can be multiplied, at least under
   certain conditions. Here, mainly motivated by some engineering  applications in the analysis of the structures,
   we propose a different definition of multiplication of
   distributions which can be easily extended to any spatial
   dimension. In particular we prove that with this new definition
   delta functions and their derivatives can still be multiplied.

\end{abstract}

\vfill

\newpage

\section{Introduction}

In the literature several examples of multiplication of
distributions exist, more or less interesting and more or less
useful for concrete applications.  In \cite{bag1} and \cite{bag2} we
have proposed our own definition of multiplication in one spatial
dimension, $d=1$, and we have proved that two or more delta
functions can be multiplied and produce, as a result, a linear
functional over $\D(\Bbb{R})$. However, our definition does not
admit a {\em natural} extension to $d>1$. This is a strong
limitation, both mathematically and for concrete applications: for
instance, it is known to the civil engineers community that the
problem of fracture mechanics may be analyzed considering beams
showing discontinuities along the beam span, \cite{cadde,cadde2}.
The {\em classical} approach for finding solutions of beams with
discontinuities consists in looking for continuous solutions between
discontinuities and imposing continuity conditions at the fractures.
In \cite{cadde,cadde2} a different strategy has been successfully
proposed, modeling the flexural stiffness and the slope
discontinuities as delta functions. However, in this approach, the
problem of multiplying two delta functions naturally arises, and
this was solved using the definition given in \cite{bag1}. This
application proved to be useful not only to get a solution in a
closed form but also to get numerical results in a reasonable simple
way, when compared with the older existing approaches. These very
promising results, however, have been discussed only in $d=1$ since
the mathematical framework, which is behind the concrete
application, only existed in one dimension, \cite{bag1,bag2}. It is
not surprising, therefore, that extensions of our procedure to
higher spatial dimensions has been strongly urged  in order to
consider the same problem for more general physical systems, i.e.
for physical systems in higher dimensions like two or
three-dimensional beams, in which the fractures can be schematized
as two or three-dimensional delta functions.

In a completely different field of research the same necessity
appeared: for instance, in the analysis of stochastic processes the
need for what is called a {\em renormalization procedure for the
powers of white noise}, where the white noise is essentially a delta
function, has produced many results, see \cite{acc}. Also,
applications to electric circuits exist which are surely less
mathematically oriented, \cite{bor}, and again are based on the
possibility of giving a meaning to $\delta^2(x)$. With these
motivations we have considered the problem of defining the
multiplication of two distributions in more than one spatial
dimension. However, our original definition cannot be easily
extended to $d>1$, because the regularization proposed in
\cite{breme} and used in \cite{bag1,bag2} does not work without
major changes in this case. For this reason we propose here a
different definition of multiplication, which works perfectly in any
spatial dimension. Moreover, this new approach returns results which
are very close to those in \cite{bag1,bag2}, for $d=1$.

The paper is organized as follows:

in the next Section we briefly recall the main definitions and
results of \cite{bag1} and \cite{bag2};

in Section 3 we propose a different definition of multiplication
in one dimension and we show that  results which are not
significantly different from  those of Section 2 are recovered;

in Section 4 we extend the definition to an arbitrary spatial
dimension and prove that with this new definition two delta
functions can be multiplied.

\section{A brief resume of our previous results}

In \cite{bag1,bag2} we have introduced a (family of) multiplications
of distributions making use of two different regularization
discussed in the literature and adapted to our purposes. Here, for
completeness' sake, we briefly recall our strategy without going
into too many details.

The first ingredient was first introduced in \cite{breme}, where it
was proven that, given a distribution $T$ with compact support, the
function \be \label{analitic}
 {\bf T}^0(z) \equiv \frac{1}{2\pi i}\, T\cdot(x-z)^{-1}
\en exists and is holomorphic in $z$ in the whole $z$-plane minus
the support of $T$. Moreover, if $T(x)$ is a continuous function
with compact support, then $T_{red}(x,\epsilon) \equiv {\bf
T}^0(x+i\epsilon)-{\bf T}^0(x-i\epsilon) $ converges uniformly to
$T(x)$ on the whole real axis for $\epsilon \rightarrow 0^+$.
Finally, if $T$ is a distribution in ${\cal D'}(\Bbb{R})$ with
compact support then $T_{red}(x,\epsilon)$ converges to $T$ in the
following sense
$$
T (\Psi) = \lim_{\epsilon \rightarrow 0} \int_{-\infty}^{\infty}
T_{red}(x,\epsilon) \, \Psi (x) \, dx
$$
for every test function $\Psi \in {\cal D}(\Bbb{R})$.

As discussed in \cite{breme}, this definition can be extended to a
larger class of one-dimensional distributions with support not
necessarily compact, ${\cal V}'(\Bbb{R})$, while it is much harder
to extend the same definition to more than one spatial dimension.

The second ingredient is the so-called  method of sequential
completion, which is discussed for instance in \cite{col}, and it
follows essentially from  very well known results on the regularity
of the convolution of distributions and test functions. Let $\Phi
\in {\cal D}(\Bbb{R})$ be a given function with supp $\Phi \subseteq
[-1,1]$ and $\int_{\Bbb{R}} \Phi (x) \, dx =1$. We call
$\delta-$sequence the sequence $\delta_n,\, n\in {\Bbb{N}},$ defined
by $\delta_n(x) \equiv n\, \Phi(nx)$. Then, $\forall \,  T \in {\cal
D'}(\Bbb{R})$, the convolution $T_n \equiv T*\delta_n$ is a
$C^{\infty}-$function for any fixed $n\in \Bbb{N}$. The sequence
$\{T_n\}$ converges to $T$ in the topology of ${\cal D'}$, when $n
\rightarrow \infty$. Moreover, if $T(x)$ is a continuous function
with compact support, then $T_n(x)$ converges uniformly to $T(x)$.

We are now ready to recall our original definition: for any couple
of distributions $T,S \, \in {\cal D'}(\Bbb{R}), \, \forall \,
\alpha, \beta
>0$ and $\forall \, \Psi \, \in {\cal D}(\Bbb{R})$ we start defining the following
quantity: \be (S\otimes T)_n^{(\alpha,\beta)}(\Psi ) \equiv
\frac{1}{2} \int_{-\infty}^{\infty} [S_n^{(\beta)}(x)\,
T_{red}(x,\frac{1}{n^\alpha}) + T_n^{(\beta)}(x)\,
S_{red}(x,\frac{1}{n^\alpha})]\, \Psi (x) \, dx  \en where \be
S_n^{(\beta)}(x) \equiv (S*\delta_n^{(\beta)})(x), \label{conv}
\en with $\delta_n^{(\beta)}(x) \equiv n^{\beta} \Phi
(n^{\beta}x)$, which is surely well defined for any choice of
$\alpha, \beta,\, T,\, S$ and $\Psi$. Moreover, if the limit of
the sequence $\left\{(S\otimes T)_n^{(\alpha,\beta)}(\Psi
)\right\}$ exists for all $\Psi(x)\in\D(\Bbb{R})$, we define
$(S\otimes T)_{(\alpha,\beta)}(\Psi )$ as: \be \label{def}
(S\otimes T)_{(\alpha,\beta)} (\Psi )\equiv \lim_{n \rightarrow
\infty} (S\otimes T)_n^{(\alpha,\beta)}(\Psi ) \en

Of course, as already remarked in \cite{bag1}, the definition
(\ref{def}) really defines  many multiplications of distributions.
In order to obtain {\underline {one definite}} product we have to
fix the positive values of $\alpha$ and $\beta$ and the particular
function $\Phi$ which is used to construct the $\delta$-sequence.
Moreover, if $T(x)$ and $S(x)$ are two continuous functions with
compact supports, and if $\alpha$ and $\beta$ are any pair of
positive real numbers, then: (i) $T_n^{(\beta)}(x)\,
S_{red}(x,\frac{1}{n^\alpha}) \mbox{ converges uniformly to }
S(x)\, T(x)$; (ii) $\forall \, \Psi(x) \in \D(\Bbb{R}) \Rightarrow
(T\otimes S)_{(\alpha,\beta)}(\Psi) = \int_{-\infty}^{\infty}
T(x)\, S(x) \, \Psi (x)\, dx$.

It is furthermore very easy to see that the product $(S\otimes T)_
{(\alpha,\beta)}$ is a linear functional on ${\cal D}(\Bbb{R})$
due to the linearity of the integral and to the properties of the
limit. The continuity of such a functional is, on the contrary,
not obvious at all, but, as formulas (\ref{resf1})-(\ref{resf6})
show, is a free benefit of our procedure.

\vspace{2mm} We now recall few results obtained in \cite{bag1,bag2}.

If we assume  $\Phi$ to be of the form \be
    \Phi(x)  = \left\{
          \begin{array}{ll}
            \frac{x^m}{N_m} \cdot \exp\{\frac{1}{x^2-1}\},    &       |x| <1 \\
           0,   &       |x| \geq 1.
       \end{array}
        \right.
\label{fi} \en where $m$ is an even natural number and $N_m$ is a
normalization constant which gives $\int_{-1}^{1} \Phi(x)\, dx=1$,
and we put $ A_j \equiv \int_{-\infty}^{\infty}
\frac{\Phi(x)}{x^j}\, dx, $ whenever it exists, we have proved that:

\noindent if $m>1$ then \be
    (\delta \otimes \delta)_{(\alpha,\beta)}  = \left\{
          \begin{array}{ll}
          \frac{1}{\pi}A_2 \delta,    &       \alpha=2\beta \\
           0,   &       \alpha>2\beta;
       \end{array}
        \right.
\label{resf1} \en if $m>2$ then \be (\delta \otimes
\delta')_{(\alpha,\beta)}  = 0 \hspace{2cm} \forall \alpha\geq
3\beta;
 \label{resf2}
\en if $m>3$ then \be
    (\delta' \otimes \delta')_{(\alpha,\beta)}  = \left\{
          \begin{array}{ll}
          \frac{-6}{\pi}A_4 \delta,    &       \alpha=4\beta \\
           0,   &       \alpha>4\beta.
       \end{array}
        \right.
\label{resf3} \en Also, if $m>3$ then \be
    (\delta \otimes \delta'')_{(\alpha,\beta)}  = \left\{
          \begin{array}{ll}
          \frac{6}{\pi}A_4 \delta,    &       \alpha=4\beta \\
           0,   &       \alpha>4\beta.
       \end{array}
        \right.
\label{resf4} \en If $m>4$ then \be (\delta' \otimes
\delta'')_{(\alpha,\beta)}  = 0 \hspace{2cm} \forall \alpha\geq
5\beta.
 \label{resf5}
\en Finally, if $m>5$ then \be
    (\delta'' \otimes \delta'')_{(\alpha,\beta)}  = \left\{
          \begin{array}{ll}
          \frac{120}{\pi}A_6 \delta,    &       \alpha=6\beta \\
           0,   &       \alpha>6\beta.
       \end{array}
        \right.
\label{resf6} \en

It is worth stressing that formula (\ref{resf1}), for
$\alpha>2\beta$, coincides with the result given by the neutrix
product discussed by Zhi and Fisher, see \cite{zhi}. Also, because
of our technique which strongly relies on the Lebesgue dominated
convergence theorem, LDCT, the above formulas only give sufficient
conditions for the product between different distributions to
exist. In other words: formula (\ref{resf1}) does not necessarily
implies that $(\delta \otimes \delta)_{(\alpha,\beta)}$ does not
exist for $\alpha<2\beta$. In this case, however, different
techniques should be used to check the existence or the
non-existence of $(\delta \otimes \delta)_{(\alpha,\beta)}$.

 More remarks on this approach can be
found in \cite{bag1} where, in particular the above results are
extended to the product between two distributions like
$\delta^{(l)}$ and $\delta^{(k)}$ for generic $l,k=0,1,2,\ldots$. In
\cite{bag2} we have further discussed the extension of the
definition of our multiplication to more distributions and to the
case of non commuting distributions, which is relevant in quantum
field theory.

\section{A different definition in $d=1$}

The idea behind the definition we introduce in this section is very
simple: since the regularization $T\rightarrow T_{red}$ cannot be
easily generalized to higher spatial dimensions, $d>1$, we use twice
the convolution procedure in (\ref{conv}), $T\rightarrow
T_n^{(\alpha)}=T\ast\delta_n^{(\alpha)}$, with different values of
$\alpha$ as we will see.

Let  $\Phi(x)\in\D(\Bbb{R})$ be a given non negative function,
with support in $[-1,1]$ and such that
$\int_{\Bbb{R}}\Phi(x)\,dx=1$. In the rest of this section, as in
\cite{bag1,bag2}, we will essentially consider the following
particular choice of $\Phi(x)$, \be
    \Phi(x)  = \left\{
          \begin{array}{ll}
            \frac{x^m}{N_m} \cdot \exp\{\frac{1}{x^2-1}\},    &       |x| <1 \\
           0,   &       |x| \geq 1.
       \end{array}
        \right.
\label{31} \en where $m$ is some fixed even natural number and
$N_m$ is a normalization constant fixed by the condition
$\int_{-1}^{1} \Phi(x)\, dx=1$. As we have discussed in
\cite{bag1} the sequence
$\delta_n^{(\alpha)}(x)=n^\alpha\Phi(n^\alpha x)$ is a {\em
delta-sequence} for any choice of $\alpha>0$. This means that: (1)
$\delta_n^{(\alpha)}(x)\rightarrow \delta(x)$ in $\D'(\Bbb{R})$
for any $\alpha>0$; (2) for any $n\in\Bbb{N}$ and for all
$\alpha>0$ if $T(x)$ is a continuous function with compact support
then the convolution
$T_n^{(\alpha)}(x)=(T\ast\delta_n^{(\alpha)})(x)$ converges to
$T(x)$ uniformly for all $\alpha>0$; (3) if $T(x)\in\D(\Bbb{R})$
then $T_n^{(\alpha)}(x)$ converges to $T(x)$ in the topology
$\tau_\D$ of $\D(\Bbb{R})$; (4) if $T\in\D'(\Bbb{R})$ then
$T_n^{(\alpha)}(x)$ is a $C^\infty$ function and it converges to
$T$ in $\D'(\Bbb{R})$ as $n$ diverges independently of $\alpha>0$.

We remark here that all these results can be naturally extended to
larger spatial dimensions, and this will be useful in the next
section.

Our next step is to replace definition (\ref{def}) with our
alternative multiplication. To begin with, let us consider two
distributions $T, S\in\D'(\Bbb{R})$, and let us compute their
convolutions $T_n^{(\alpha)}(x)=(T\ast\delta_n^{(\alpha)})(x)$ and
$S_n^{(\beta)}(x)=(T\ast\delta_n^{(\beta)})(x)$ with
$\delta_n^{(\alpha)}(x)=n^\alpha\Phi(n^\alpha x)$, for
$\alpha,\beta>0$ to be fixed in the following. $T_n^{(\alpha)}(x)$
and $S_n^{(\beta)}(x)$ are both $C^\infty$ functions, so that for
any $\Psi(x)\in\D(\Bbb{R})$ and for each fixed $n\in\Bbb{N}$, the
following integrals surely exist: \be (S\odot
T)_n^{(\alpha,\beta)}(\Psi ) \equiv \frac{1}{2} \int_{\Bbb{R}}
\left[S_n^{(\alpha)}(x)\, T_n^{(\beta)}(x) +
S_n^{(\beta)}(x)\,T_n^{(\alpha)}(x) \right]\, \Psi (x) \, dx,
\label{32}\en \be (S\odot_d T)_n^{(\alpha,\beta)}(\Psi ) \equiv
\int_{\Bbb{R}} S_n^{(\alpha)}(x)\, T_n^{(\beta)}(x) \, \Psi (x) \,
dx, \label{33}\en \be (S\odot_{ex} T)_n^{(\alpha,\beta)}(\Psi )
\equiv \int_{\Bbb{R}} S_n^{(\beta)}(x)\, T_n^{(\alpha)}(x) \, \Psi
(x) \, dx. \label{34}\en The suffix $d$ stands for {\em direct},
while $ex$ stands for {\em exchange}. This is because the two
related integrals remind us the direct and the exchange
contributions in an energy Hartree-Fock computation, typical of
quantum many-body systems. It is clear that if $S\equiv T$ then the
three integrals above coincide: $(S\odot S)_n^{(\alpha,\beta)}(\Psi
)=(S\odot_d S)_n^{(\alpha,\beta)}(\Psi )=(S\odot_{ex}
S)_n^{(\alpha,\beta)}(\Psi )$, for all $\alpha,\beta,\Psi,n$. In
general, however, they are different and we will discuss an example
in which they really produce different results when
$n\rightarrow\infty$. We say that the distributions $S$ and $T$ are
$\odot$-multiplicable, and we indicate with $(S\odot
T)_{(\alpha,\beta)}$ their product, if the following limit exists
for all $\Psi(x)\in\D(\Bbb{R})$: \be (S\odot
T)_{(\alpha,\beta)}(\Psi )=\lim_{n,\,\infty}(S\odot
T)_n^{(\alpha,\beta)}(\Psi ).\label{35}\en Analogously we put, when
the following limits exist, \be (S\odot_d T)_{(\alpha,\beta)}(\Psi
)=\lim_{n\,\rightarrow\,\infty}(S\odot_d T)_n^{(\alpha,\beta)}(\Psi
).\label{36}\en and \be (S\odot_{ex} T)_{(\alpha,\beta)}(\Psi
)=\lim_{n\,\rightarrow\,\infty}(S\odot_{ex}
T)_n^{(\alpha,\beta)}(\Psi ).\label{37}\en

Again, it is clear that, whenever they exist, $(S\odot
S)_{(\alpha,\beta)}(\Psi )=(S\odot S)_{(\beta,\alpha)}(\Psi
)=(S\odot_d S)_{(\alpha,\beta)}(\Psi )=(S\odot_{ex}
S)_{(\alpha,\beta)}(\Psi )$, for all $\Psi(x)\in\D(\Bbb{R})$ while
they are different, in general, if $S\neq T$. Of course, the
existence of these limits in general will depend on the values of
$\alpha$ and $\beta$ and on the particular choice of $\Phi(x)$. For
this reason, as in \cite{bag1,bag2}, we are really defining a {\em
class of multiplications} of distributions and not just a single
one. We have already discussed in \cite{bag1} a simple example which
shows how these different multiplications may have a physical
interpretation in a simple quantum mechanical system. We will
discuss further physical applications of our procedure in a
forthcoming paper.

We now list a set of properties which follow directly from the
definitions:
\begin{enumerate}
\item if $S(x)$ and $T(x)$ are continuous functions with compact
support then \be (S\odot T)_{(\alpha,\beta)}(\Psi )=(S\odot_d
T)_{(\alpha,\beta)}(\Psi )=(S\odot_{ex} T)_{(\alpha,\beta)}(\Psi
)=\int_{\Bbb{R}}\,S(x)\,T(x)\,\Psi(x)\,dx, \label{38}\en for all
$\Psi(x)\in\D(\Bbb{R})$ and for all $\alpha,\beta>0$.
\item for all fixed $n$, for all $\alpha,\beta>0$ and for all
$\Psi(x)\in\D(\Bbb{R})$ we have \be (S\odot_d
T)_n^{(\alpha,\beta)}(\Psi )=(S\odot_{ex} T)_n^{(\beta,\alpha)}(\Psi
)\label{39}\en
\item for all $n$, $\alpha, \beta>0$ and for all $\Psi(x)\in\D(\Bbb{R})$
we have \be \left(S\odot_d
T\right)_n^{(\alpha,\beta)}(\Psi)=\left(T\odot_d
S\right)_n^{(\beta,\alpha)}(\Psi), \mbox{ and } \left(S\odot_{ex}
T\right)_n^{(\alpha,\beta)}(\Psi)=\left(T\odot_{ex}
S\right)_n^{(\beta,\alpha)}(\Psi) \label{310}\en

\item given $S,T\in\D'(\Bbb{R})$ such that the following quantities exist we
have \be (S\odot T)_{(\alpha,\beta)}(\Psi
)=\frac{1}{2}\left\{(S\odot_d T)_{(\alpha,\beta)}(\Psi
)+(S\odot_{ex} T)_{(\alpha,\beta)}(\Psi )\right\}\label{311}\en

\end{enumerate}

We will now discuss a few examples of these definitions, beginning
with  maybe the most relevant for concrete applications, i.e. the
multiplication of two delta functions.

\vspace{2mm}

{\bf Example nr.1: $\left(\delta\odot
\delta\right)_{(\alpha,\beta)}$ }

\vspace{2mm}

First of all we remind that, in this case, the $\odot$, $\odot_d$
and $\odot_{ex}$ multiplications all coincide. If the following
limit exists for some $\alpha, \beta>0$, we have
$$
\left(\delta\odot
\delta\right)_{(\alpha,\beta)}(\Psi)=\left(\delta\odot
\delta\right)_{(\beta,\alpha)}(\Psi)=\lim_{n\,\rightarrow\,\infty}\int_{\Bbb{R}}
\delta_n^{(\alpha)}(x)\, \delta_n^{(\beta)}(x) \, \Psi (x) \, dx=$$
$$=\lim_{n\,\rightarrow\,\infty}n^{\alpha+\beta}\int_{\Bbb{R}} \Phi(n^\alpha\,
x)\, \Phi(n^\beta\, x)\, \, \Psi (x) \, dx,
$$
for all $\Psi(x)\in\D(\Bbb{R})$. It is an easy exercise to check
that this limit does not exist, if $\Psi(0)\neq0$ and
$\alpha=\beta$. Therefore we consider, first of all, the case
$\alpha>\beta$. In this case we can write
$$
\left(\delta\odot
\delta\right)_n^{(\alpha,\beta)}(\Psi)=n^{\beta}\,\int_{-1}^1\,\Phi(x)\,\Phi(xn^{\beta-\alpha})\,
\Psi(xn^{-\alpha})\,dx,
$$
and the existence of its limit can be proved, as in
\cite{bag1,bag2}, using the LDCT. Choosing $\Phi(x)$ as in
(\ref{31}), and defining
$B_m=\frac{1}{eN_m}\,\int_{-1}^1\,x^m\,\Phi(x)\,dx$, which is
surely well defined and strictly positive  for all fixed even $m$,
it is quite easy to deduce that \be
 (\delta \odot \delta)_{(\alpha,\beta)}(\Psi)  = \left\{
          \begin{array}{ll}
          B_m \Psi(0)= B_m\,\delta(\Psi),    &       \alpha=\beta\left(1+\frac{1}{m}\right) \\
           0,   &       \alpha>\beta\left(1+\frac{1}{m}\right).
       \end{array}
        \right.
\label{312} \en  As we can see, this result is quite close to the
one in (\ref{resf1}). Analogously to what already stressed in
Section 2, formula (\ref{312}) does not imply that $(\delta \odot
\delta)_{(\alpha,\beta)}(\Psi)$ does not exist if
$\alpha<\beta\left(1+\frac{1}{m}\right)$ because using the LDCT we
only find sufficient conditions for $(\delta \odot
\delta)_{(\alpha,\beta)}(\Psi)$ to exist. However, with respect to
our results in \cite{bag1}, here we can say a bit more, because we
have $(\delta \odot \delta)_{(\alpha,\beta)}(\Psi)=(\delta \odot
\delta)_{(\beta,\alpha)}(\Psi)$, which was not true in general for
the multiplication $\otimes_{(\alpha,\beta)}$ introduced in
\cite{bag1}. Therefore we find that \be
 (\delta \odot \delta)_{(\alpha,\beta)}(\Psi)  = \left\{
          \begin{array}{ll}
           B_m\,\delta(\Psi),    &       \alpha=\beta\left(1+\frac{1}{m}\right)^{-1},\,\mbox{ or }
           \alpha=\beta\left(1+\frac{1}{m}\right) \\
           0,   &       \alpha<\beta\left(1+\frac{1}{m}\right)^{-1} \,
           \hspace{2mm} \mbox{ or } \alpha>\beta\left(1+\frac{1}{m}\right),
       \end{array}
        \right.
\label{313} \en while nothing can be said in general if
$\alpha\in\left]\beta\left(1+\frac{1}{m}\right)^{-1},\beta\left(1+\frac{1}{m}\right)\right[$.

\vspace{2mm}

{\bf Example nr.2: $\left(\delta\odot
\delta'\right)_{(\alpha,\beta)}$ }

\vspace{2mm}

In this case the three multiplications $\odot$, $\odot_d$ and
$\odot_{ex}$ do not need to coincide. Indeed we will find serious
differences between the three, as expected.

First of all we concentrate on the computation of
$\left(\delta\odot_d \delta'\right)_{(\alpha,\beta)}$. Because of
(\ref{39}) this will also produce $\left(\delta\odot_{ex}
\delta'\right)_{(\beta,\alpha)}$. We have
$$
\left(\delta\odot_d
\delta'\right)_n^{(\alpha,\beta)}(\Psi)=n^{\alpha+2\beta}\,\int_{\Bbb{R}}\,\Phi(n^\alpha\,x)\,\Phi'(n^\beta\,x)\,
\Psi(x)\,dx,
$$
where $\Phi'$ is the derivative of $\Phi$. Again, it is easy to
check that the limit of this sequence does not exist, if
$\alpha=\beta$, for all $\Psi(x)\in\D(\Bbb{R})$ but only for those
$\Psi(x)$ which go to zero fast enough when $x\rightarrow 0$. Let us
then consider $\left(\delta\odot_d
\delta'\right)_n^{(\alpha,\beta)}(\Psi)$ for $\alpha>\beta$. In this
case we can write
$$
\left(\delta\odot_d
\delta\right)_n^{(\alpha,\beta)}(\Psi)=n^{2\beta}\,\int_{-1}^1\,\Phi(x)\,\Phi'(xn^{\beta-\alpha})\,
\Psi(xn^{-\alpha})\,dx,
$$
and by the LDCT we deduce that \be
 (\delta \odot_d \delta')_{(\alpha,\beta)}(\Psi)  = \left\{
          \begin{array}{ll}
          K_m \,\delta(\Psi),    &       \alpha=\beta\,\frac{m+1}{m-1} \\
           0,   &       \alpha>\beta\,\frac{m+1}{m-1},
       \end{array}
        \right.
\label{315} \en where
$K_m=\frac{m}{N_me}\,\int_{-1}^1\,x^{m-1}\,\Phi(x)\,dx$. We see
that, contrarily to (\ref{resf2}), we can obtain a non trivial
result with the $\odot_d$ multiplication. It is clear therefore
that also the $(\delta \odot_{ex}\delta')$ can be non trivial, as
remarked above .

The situation is completely different for the $(\delta \odot
\delta')_{(\beta,\alpha)}(\Psi)$. Indeed, it is not difficult to
understand that the LDCT cannot be used to prove its existence. The
reason is quite general and is the following:

suppose that for two distributions $T$ and $S$ in $\D'(\Bbb{R})$ $(T
\odot_d S)_{(\alpha,\beta)}$ exists for $\alpha$ and $\beta$ such
that $\alpha>\gamma\beta$, where $\gamma$ is some constant larger
than 1 appearing because of the LDCT. For instance here
$\gamma=\frac{m+1}{m-1}$, while in Example nr.1 we had
$\gamma=1+\frac{1}{m}$. Therefore, since if $(S\odot_{ex}
T)_{(\beta,\alpha)}$ and $(S\odot_d T)_{(\alpha,\beta)}$ both exist
and coincide, using (\ref{311}) we have \be (S\odot
T)_{(\alpha,\beta)}(\Psi )=\frac{1}{2}\left\{(S\odot_d
T)_{(\alpha,\beta)}(\Psi )+(S\odot_{d} T)_{(\beta,\alpha)}(\Psi
)\right\}\label{316}\en Of course, $(S\odot T)_{(\alpha,\beta)}(\Psi
)$ exists if $(S\odot_{d} T)_{(\alpha,\beta)}(\Psi )$ and
$(S\odot_{d} T)_{(\beta,\alpha)}(\Psi )$ both exist, which in turn
implies that $\alpha>\gamma\beta$ and, at the same time, that
$\beta>\gamma\alpha$. These are clearly incompatible. Therefore in
order to check whether $(S\odot T)_{(\alpha,\beta)}(\Psi )$ exists
or not it is impossible to use the LDCT which only gives sufficient
conditions for the multiplication to be defined: some different
strategy should be considered.

\vspace{2mm}

{\bf Example nr.3: $\left(\delta'\odot
\delta'\right)_{(\alpha,\beta)}$ }

\vspace{2mm}

As for Example nr.1 we remark that here the $\odot$, $\odot_d$ and
$\odot_{ex}$ multiplications all coincide. If the following limit
exists for some $\alpha, \beta>0$, we have
$$
\left(\delta'\odot
\delta'\right)_{(\alpha,\beta)}(\Psi)=\left(\delta'\odot
\delta'\right)_{(\beta,\alpha)}(\Psi)=\lim_{n\,\rightarrow\,\infty}\int_{\Bbb{R}}
\delta_n^{\,'(\alpha)}(x)\, \delta_n^{\,'(\beta)}(x) \, \Psi (x) \,
dx=$$
$$=\lim_{n\,\rightarrow\,\infty}n^{2\alpha+2\beta}\int_{\Bbb{R}} \Phi'(n^\alpha\,
x)\, \Phi'(n^\beta\, x)\, \, \Psi (x) \, dx,
$$
for all $\Psi(x)\in\D(\Bbb{R})$. As before, it is quite easy  to
check that this limit does not exist, if $\Psi(0)\neq0$, when
$\alpha=\beta$. Therefore we start considering the case
$\alpha>\beta$. In this case we can write
$$
\left(\delta'\odot
\delta'\right)_n^{(\alpha,\beta)}(\Psi)=n^{\alpha+2\beta}\,\int_{-1}^1\,\Phi'(x)\,\Phi'(xn^{\beta-\alpha})\,
\Psi(xn^{-\alpha})\,dx,
$$
and again the existence of its limit can be proved using the
 LDCT. Choosing $\Phi(x)$ as
in (\ref{31}) and defining $\tilde
B_m=\frac{m}{eN_m}\,\int_{-1}^1\,x^{m-1}\,\Phi'(x)\,dx$, which
surely exists for all fixed even $m$, we deduce that \be
 (\delta' \odot \delta')_{(\alpha,\beta)}(\Psi)  = \left\{
          \begin{array}{ll}
          \tilde B_m \,\delta(\Psi),    &       \alpha=\beta\,\frac{m+1}{m-2} \\
           0,   &       \alpha>\beta\,\frac{m+1}{m-2}.
       \end{array}
        \right.
\label{316b} \en  However this is not the end of the story,
because we still can use the symmetry $(\delta' \odot
\delta')_{(\alpha,\beta)}(\Psi)=(\delta' \odot
\delta')_{(\beta,\alpha)}(\Psi)$ discussed before. We find \be
 (\delta' \odot \delta')_{(\alpha,\beta)}(\Psi)  = \left\{
          \begin{array}{ll}
           \tilde B_m\,\delta(\Psi),    &       \alpha=\beta\,\frac{m-2}{m+1},\,\mbox{ or }
           \alpha=\beta\,\frac{m+1}{m-2} \\
           0,   &       \alpha<\beta\,\frac{m-2}{m+1},\,\mbox{ or }
           \alpha>\beta\,\frac{m+1}{m-2},
       \end{array}
        \right.
\label{317} \en while nothing can be said in general if
$\alpha\in\left]\beta\,\frac{m-2}{m+1},\beta\,\beta\,\frac{m+1}{m-2}\right[$.
Needless to say, we need here to restrict to the following values of
$m$: $m=4,6,8,\ldots$.

\vspace{2mm}

Summarizing we find that results which are very close to those in
\cite{bag1} can be recovered with each one of the definitions in
(\ref{35}), (\ref{36}) or (\ref{37}). The main differences
essentially arise from the lack of symmetries of these  two last
definitions compared to definition (\ref{35}) and the one in
\cite{bag1}.

\section{More spatial dimensions and conclusions}

While the definition of the multiplication given in \cite{bag1}, as
we have stressed before, cannot be extended easily to ${\Bbb{R}}^d$,
$d>1$, definitions (\ref{35}), (\ref{36}) or (\ref{37}) admit a
natural extension to any spatial dimensions. We concentrate here
only on the symmetric definition, (\ref{35}), since it is the most
relevant one for the application we are interested in here. Of
course no particular differences appear in the attempt of extending
$\odot_d$ and $\odot_{ex}$ to $d>1$.

The starting point is again a given non negative function
$\Phi(\underline{x})\in\D(\Bbb{R}^d)$ with support in
$I_1:=\underbrace{[-1,1]\times\cdots\times[-1,1]}_{d \mbox{
times}}$, and such that
$\int_{I_1}\Phi(\underline{x})\,d\underline{x}=1$. In this case the
delta-sequence is
$\delta_n^{(\alpha)}(\underline{x})=n^{d\alpha}\Phi(n^\alpha
\underline{x})$, for any choice of $\alpha>0$. The same results
listed in the previous section again hold in this more general
situation. For instance, if $T\in\D'({\Bbb{R}}^d)$ then
$\{T_n^{(\alpha)}(\underline{x})=(T\ast\delta_n^{(\alpha)})(\underline{x})\}$
is a sequence of $C^\infty$ functions and it converges to $T$ in
$\D'({\Bbb{R}}^d)$ as $n$ diverges, independently of $\alpha>0$.

Therefore, let us consider two distributions $T,
S\in\D'(\Bbb{R}^d)$, and let us consider their convolutions
$T_n^{(\alpha)}(\underline{x})=(T\ast\delta_n^{(\alpha)})(\underline{x})$
and
$S_n^{(\beta)}(\underline{x})=(S\ast\delta_n^{(\beta)})(\underline{x})$
with $\delta_n^{(\alpha)}(\underline{x})=n^{d\alpha}\Phi(n^\alpha
\underline{x})$, for $\alpha,\beta>0$. As usual,
$T_n^{(\alpha)}(\underline{x})$ and $S_n^{(\beta)}(\underline{x})$
are both $C^\infty$ functions, so that  the following integral
surely exists: \be (S\odot T)_n^{(\alpha,\beta)}(\Psi ) \equiv
\frac{1}{2} \int_{\Bbb{R}^d} \left[S_n^{(\alpha)}(\underline{x})\,
T_n^{(\beta)}(\underline{x}) +
S_n^{(\beta)}(\underline{x})\,T_n^{(\alpha)}(\underline{x})
\right]\, \Psi (\underline{x}) \, d\underline{x}, \label{41}\en
$\forall\,\Psi(\underline{x})\in\D(\Bbb{R}^d)$. As before the two
distributions $S$ and $T$ are $\odot$-multiplicable if the following
limit exists independently of $\Psi(\underline{x})\in\D(\Bbb{R}^d)$:
\be (S\odot T)_{(\alpha,\beta)}(\Psi
)=\lim_{n\,\rightarrow\,\infty}(S\odot T)_n^{(\alpha,\beta)}(\Psi
).\label{42}\en

We consider in the following the $\odot$-multiplication of two
delta functions, considering two different choices for the
function $\Phi(\underline{x})$ both extending the one-dimensional
case.

The starting point of our computation is the usual one: if it
exists, $(\delta\odot \delta)_{(\alpha,\beta)}(\Psi )$ must be such
that
$$
(\delta\odot \delta)_{(\alpha,\beta)}(\Psi
)=\lim_{n\,\rightarrow\,\infty}\,n^{d\alpha+d\beta}\int_{\Bbb{R}^d}
\Phi(n^\alpha\, \underline{x})\, \Phi(n^\beta\, \underline{x})\, \,
\Psi (\underline{x}) \, d\underline{x}.
$$
Again, this limit does not exist if $\alpha=\beta$, but for very
peculiar functions $\Psi(\underline{x})$. If we consider what
happens for $\alpha>\beta$ then the limit exists under certain extra
conditions.

For instance, if we take
$\Phi(\underline{x})=\prod_{j=1}^d\,\Phi(x_j)$, where $\Phi(x_j)$ is
the one in (\ref{31}), the computation factorizes and the final
result, considering also the symmetry of the multiplication, is a
simple extension of the one in (\ref{313}): \be
 (\delta \odot \delta)_{(\alpha,\beta)}(\Psi)  = \left\{
          \begin{array}{ll}
           B_m^d\,\delta(\Psi),    &       \alpha=\beta\left(1+\frac{1}{m}\right)^{-1},\,\mbox{ or }
           \alpha=\beta\left(1+\frac{1}{m}\right) \\
           0,   &       \alpha<\beta\left(1+\frac{1}{m}\right)^{-1}
           \hspace{2mm}
           \mbox{ or }  \alpha>\beta\left(1+\frac{1}{m}\right),
       \end{array}
        \right.
\label{43} \en \vspace{2mm} A different choice of
$\Phi(\underline{x})$, again related to the one in (\ref{31}), is
the following: \be
    \Phi(\underline{x})  = \left\{
          \begin{array}{ll}
            \frac{\|\underline{x}\|^m}{N'_m} \, \exp\{\frac{1}{\|\underline{x}\|^2-1}\},    &       \|\underline{x}\| <1 \\
           0,   &       \|\underline{x}\| \geq 1,
       \end{array}
        \right.
\label{44} \en where $N'_m$ is a normalization constant and
$\|\underline{x}\|=\sqrt{x_1^2+\cdots +x_d^2}$. With this choice,
defining
$C_m=\frac{1}{N'_me}\int_{\Bbb{R}}\|\underline{x}\|\,\Phi(\underline{x})\,d\underline{x}$
and using the symmetry property of $\odot_{(\alpha,\beta)}$, we find
\be
 (\delta \odot \delta)_{(\alpha,\beta)}(\Psi)  = \left\{
          \begin{array}{ll}
           C_m\,\delta(\Psi),    &       \alpha=\beta\left(1+\frac{d}{m}\right)^{-1},\,\mbox{ or }
           \alpha=\beta\left(1+\frac{d}{m}\right) \\
           0,   &       \alpha<\beta\left(1+\frac{d}{m}\right)^{-1}
           \hspace{2.5mm}
           \mbox{ or }  \alpha>\beta\left(1+\frac{d}{m}\right).
       \end{array}
        \right.
\label{45} \en Therefore the limit defining the product of two delta
can be defined, at least under certain conditions, also with this
choice of $\Phi(\underline{x})$ . The main differences between the
above results are the values of the constants and the fact that $d$
explicitly appears  in the result in (\ref{43}), while it only
appears in the condition relating $\alpha$ and $\beta$ in
(\ref{45}).

\vspace{3mm}

The conclusion of this short paper is that the use of sequential
completion, properly adapted for our interests, is much simpler
\underline{and} convenient. The next step of our analysis will be
 to use our results in applications to three-dimensional
engineering structures, trying to extend the results in
\cite{cadde,cadde2}.

\vspace{50pt}

\noindent{\large \bf Acknowledgments} \vspace{5mm}

This work has been partially supported by M.U.R.S.T.

\end{document}